\documentclass{emulateapj}
\RequirePackage{lineno}
\usepackage{amssymb}
\usepackage{amsmath}
\usepackage{mathrsfs}
\usepackage{natbib}
\usepackage{array}
\usepackage{float}
\usepackage{graphicx}
\usepackage{subfigure}
\usepackage{color}
\usepackage[toc,page]{appendix}

\def\beq{\begin{equation}}
\def\enq{\end{equation}}
\def\bea{\begin{eqnarray}}
\def\ena{\end{eqnarray}}

\def\jcap{Jour. Cosmology and Astro-Particle Phys.\,}

\begin{document}

\title{High-Energy Neutrino Emission from White Dwarf Mergers}
\author{Di Xiao\altaffilmark{1,2,3}, Peter M\'esz\'aros\altaffilmark{3}, Kohta Murase\altaffilmark{3}, and Zi-Gao Dai\altaffilmark{1,2}}
\affil{\altaffilmark{1}School of Astronomy and Space Science, Nanjing University, Nanjing 210093, China}
\affil{\altaffilmark{2}Key Laboratory of Modern Astronomy and Astrophysics (Nanjing University), Ministry of Education, China}
\affil{\altaffilmark{3}Department of Physics; Department of Astronomy and Astrophysics; Center for Particle and Gravitational Astrophysics, The Pennsylvania State University, University Park, PA 16802, USA}

\date{\today}

\begin{abstract}
The merger of two white dwarfs is expected to result in a central fast rotating core surrounded by a 
debris disk, in which magnetorotational instabilities give rise to a hot magnetized corona and a 
magnetized outflow. The dissipation of magnetic energy via reconnection could lead to the acceleration 
of cosmic-rays in the expanding material, which would result in high energy neutrinos.
We discuss the possibility of using these neutrino signals as probes of the outflow dynamics, 
magnetic energy dissipation rate and cosmic-ray acceleration efficiency. Importantly, the accompanying high-energy gamma-rays are absorbed within these sources because of the 
large optical depth, so these neutrino sources can be regarded as {\it hidden cosmic-ray accelerators} 
that are consistent with the non-detection of gamma-rays with {\it Fermi-LAT}. 
While the cosmic-ray generation rate is highly uncertain, if it reaches $\sim10^{45}\,\rm 
erg\,Mpc^{-3}\,yr^{-1}$, the diffuse neutrino flux could contribute a substantial fraction of 
the IceCube observations. We also evaluate the prospect of 
observing individual merger events, which provides a means for testing such sources in the future.    
\end{abstract}


\section{Introduction}

After both stars in a binary system have ended their main sequence phase, a common result is a
white dwarf (WD) binary, and a large fraction of these binaries are expected to merge in less than a 
Hubble time \citep{nel01}, due to various causes like gravitational wave emission or magnetic braking 
and so on. Numerical simulations indicate that the post-merger system consists of a fast-rotating central 
core surrounded by a Keplerian disk \citep{ji13}, which have inherited the orbital angular momentum of 
the progenitor WD binary. A hot corona above the disk can form as a result of the development of the
magnetorotational instability within the disk. This corona is highly magnetized, with field strengths 
of order $10^{10}-10^{11}\,\rm G$ within a radius of $R\sim 10^9\,\rm cm$, with strong outflows  emerging from this central region (see Figure 2 in \citet[][]{ji13}). The ejecta velocity of this outflow is of order $10^9\,\rm cm\,s^{-1}$ and the total ejected mass is about $10^{-3}~M_\odot\sim10^{30}\,\rm g$, creating an expanding fireball. \citet{bel14} argued that internal shocks may develop in this outflowing fireball, increasing its radiative output and leading to a predicted bright optical transient of luminosity $10^{41}-10^{42}\,\rm erg\,s^{-1}$. 
Here we examine possible energy dissipation scenarios in such merger events which could lead to cosmic-ray (CR) acceleration and high-energy neutrino emission. 

The origin of the diffuse high-energy (TeV to PeV) neutrino flux discovered by IceCube 
\citep{aar13a,aar13b,aar15,IC3+15tevnu,aar15c,aar16} is currently under intense debate. 
Among various possible astrophysical sources which can contribute to this flux \citep[e.g., for reviews, see][]{wax13,mes14,anc14,Ahlers+15nurev}, the most commonly discussed are CR reservoirs including star-forming galaxies (SFGs) and starbursts galaxies (SBGs) \citep{loe06}, and galaxy clusters and groups \citep{mur08,kot09}, 
in which confined CRs can produce neutrinos via $pp$ interactions. 
In SFGs and SBGs, CRs are accelerated in supernova and hypernova remnants~\citep{mur13,sov15,nick15, cha15,xia16} as well as fast outflows and possible jets from active galactic nuclei~\citep{mur13, tam14, mur14,wl16}. 
The IceCube neutrino flux can be accounted for in the SBG scenario without violating the extragalactic gamma-ray background \citep{mur13,cha14,xia16} or even the simultaneous explanation of neutrinos, gamma-rays and CRs is possible \citep{mw16}.  
However, the neutrino data point around $30\,\rm TeV$ remains unsolved, and the fact that a large fraction of the isotropic extragalactic gamma-ray background can be explained by other sources such as blazars serves as motivation for investigating ``hidden" (i.e. $\gamma$-ray dim) neutrino sources \citep{mur16a}. 
In this work, we argue that WD merger events belong to such kinds of hidden CR accelerators, as often discussed in the context of low-power gamma-ray bursts and choked jets~\citep{mw01,mi13,xia14,xia15,senno16,ta16}. 
We calculate neutrino spectra of individual mergers and show that nearby events are detectable if CRs are accelerated efficiently by magnetic dissipation. We also discuss the diffuse neutrino flux, and find that WD mergers could also provide an interesting fraction of the IceCube diffuse neutrino background.

The paper is organized as follows. We introduce the model and method of calculation in Section 2. Then we consider the possibility of detection of individual WD merger sources in Section 3.
Section 4 presents our results of the diffuse neutrino flux from WD mergers, compared with the 
IceCube data. Lastly, we provide a summary and final discussion in Section 5.

\section{Model and Calculation}
\subsection{Outflow from the merger region}
\label{sec:outflow}
The merger of a white dwarf binary is expected to generate strong magnetic fields ($10^{10}-10^{11}$ G), as demonstrated by recent numerical simulations \citep[e.g.][]{ji13, zhu15}. For various types of WD, the birth and merger rate is slightly different \citep{bog09}. Here in our work we consider typical carbon-oxygen WD binary of equal mass $\sim0.6\,M_\odot$. 
Consistent with these simulations, to within an order of magnitude, the outflow from the central core and disrupted disk plus corona region can be characterized by an initial radius $R_0\sim 10^9 R_9~\,\rm cm$, magnetic field $B_0\sim 10^{10}B_{10}\,\rm G$, temperature $T_0\sim 10^8 T_8\,\rm K$, and mass outflow rate ${\dot M} \sim 2\times 10^{26}\,\rm g\,s^{-1}$. Considering a residence time for the baryons inside this region of order or somewhat longer than the sound crossing or virial time, $t_c \sim 100~\rm s$, the mass density at the boundary is $\rho_0\sim {\dot M}t_c/(4\pi R_0^3) \sim 1~{\dot M}_{26} (t_c/100 ~\rm s)$ g cm$^{-3}$, and the outflow velocity is $v_0\sim 10^9 \,\rm cm\,s^{-1}$, \citep[e.g.,][]{ji13,bel14}.  This magnetic field is characteristic of the corona above the disk, and the total instantaneous coronal magnetic energy is 
$\mathcal{E}_{B,0}\sim 10^{48}\,\rm erg$. One can verify that for these conditions, the initial magnetic energy density $B_0^2/8\pi\sim 4\times 10^{18}~B_{10}^2~\rm erg~cm^{-3}$, initial kinetic energy density $(1/2)\rho_0 v_0^2\sim 0.5\times 10^{18}~\rho_0 v_9^2~\rm erg~cm^{-3}$ and initial radiation energy density $a_R T_0^4\sim 0.75\times 10^{18}~T_8^4~\rm erg~cm^{-3}$ are in approximate equipartition. However, it is possible to have parameters that are not covered by the present numerical simulations.

The magnetic luminosity of the outflow can be estimated as $L_B\sim \mathcal{E}_B/t_c 
\sim 4\pi R_0^2 v_0(B_0^2/8\pi) \sim 10^{46}~\rm erg/s$, which may be an optimistic maximum luminosity.
A more conservative estimate \citep{bel14} is $L_B\sim 10^{44}~\rm erg/s$.
The outflow is expected to last for a viscous time $t_{\rm visc}$ characterizing the draining of the 
disrupted disk, which in terms of an $\alpha_B$ magnetic viscosity prescription is $t_{\rm visc}\sim 
\alpha_B^{-1}(H/r)^{-2}\Omega^{-1}\simeq 3\times10^3\,\rm s$, where we assume $\alpha_B\sim 10^{-3}$ \citep[e.g.][]{bel13, hos13} and the scale height at the tidal disruption radius $H/r\sim 1$ \citep{kas15}. As a result of the merger, the total amount
of magnetic energy ejected by the remnant is $\mathcal{E}_B\sim L_B t_{\rm visc}\sim 10^{48}-10^{50}~\rm erg$.

The toroidal component of the magnetic field in the outflow would decrease as $B_t\propto R^{-1}$, as 
in a pulsar striped wind or magnetar wind, while a poloidal component decreases as $B_p\propto R^{-2}$. In general, we can expect the total magnetic field to scale as a power-law in radius, and the magnetic 
luminosity to scale as a power-law in radius, $L_B\sim 4\pi R^2 v (B^2/8\pi) \propto R^{\alpha}$. 
For random equipartition magnetic fields, acting as a relativistic gas, purely adiabatic expansion 
would imply $L_B\propto R^{-2/3}$, i.e. $\alpha=-2/3$. Alternatively, for a strong ordered magnetic field, 
one would have $B\propto R^{-1}$ and $L_B=\rm const$, i.e. $\alpha=0$. This latter case is of  
greater interest for our purposes here.

The evolution of other quantities in the outflow are calculated as follows \citep[e.g.,][]{mur16b}. Since the outflow ceases at $t_{\rm visc}\sim3\times10^3\,\rm s$, at time $t<t_{\rm visc}$, we get $\rho\propto R^{-2}$ and after that $\rho\propto R^{-3}$.
We can define a critical radius $R_{\rm cr}\equiv v_0t_{\rm visc}=3\times10^{12}\,\rm cm$, so
\beq
\rho(R)=
	\begin{cases}
    \rho_0(R/R_0)^{-2} &\mbox{if $R_0\leq R<R_{\rm cr}$,}\\
    \rho_0(R_{\rm cr}/R_0)^{-2}(R/R_{\rm cr})^{-3} &\mbox{if \,\,\,\,\,\,$R\geq R_{\rm cr}$.}
\end{cases}
\label{eq:rho}
\enq

The blackbody temperature evolution follows $T\propto \rho^{1/3}$ for adiabatic index $4/3$, thus
\beq
T(R)=
	\begin{cases}
    T_0(R/R_0)^{-2/3} &\mbox{if $R_0\leq R<R_{\rm cr},$}\\
    T_0(R_{\rm cr}/R_0)^{-2/3}(R/R_{\rm cr})^{-1} &\mbox{if \,\,\,\,\,\,$R\geq R_{\rm cr}$.}
\end{cases}
\label{eq:temp}
\enq

\subsection{Diffusion radius and magnetic energy dissipation}
\label{sec:diss}

Initially the outflow is optical thick $\tau_{T,0}\gg 1$, and the radiation is thermalized and trapped. As the 
expansion proceeds, the optical depth decreases and when the characteristic diffusion length is equal to the characteristic dimension of the outflow (e.g. the radius of the leading gas particles), the radiation 
begins to diffuse out faster then the gas expands. The diffusion timescale is $t_{\rm diff}\sim \tau_TR/c$, where the Thomson depth evolves as
\beq
\tau_T(R)=
	\begin{cases}
    \tau_{T,0}(R/R_0)^{-1} &\mbox{if $R_0\leq R<R_{\rm cr}$,}\\
    \tau_{T,0}(R_{\rm cr}/R_0)^{-1}(R/R_{\rm cr})^{-2} &\mbox{if \,\,\,\,\,\,$R\geq R_{\rm cr}$.}
\end{cases}
\label{eq:tauT}
\enq
The expansion time is $t_{\rm exp}\sim R/v_0$, and by equaling $t_{\rm diff}=t_{\rm exp}$ we can get the diffusion radius of $R_D=6.3\times10^{13}\,\rm cm$, which naturally falls into $R>R_{\rm cr}$ regime.

For radii less than the diffusion radius, $R \leq R_D$, it is likely that any magnetic reconnection
process is suppressed, due to the high photon drag. In this regime, radiation pressure works against 
the development of turbulence and against regions of opposite magnetic polarity approaching. Beyond
$R_D$, however, radiation pressure start to drop, and reconnection may start to occur, although the
transition threshold from one regime to the other is not well-known \citep[][and references therein]{Uzdensky11magrec}.
The Thomson optical depth at the diffusion radius $\tau_T(R_D) \sim c/v$ remains above unity for at least two orders of magnitude in radius beyond $R_D$, and the dependence of the magnetic reconnection rate on the flow parameters in this still optically thick regime is speculative. For simplicity, here we adopt a simple power-law dependence for the magnetic energy dissipation rate, $\dot{\mathcal{E}}_B\propto t^{-q}$, where $q$ is a phenomenological parameter. Thus, beyond the dissipation radius, for $v=$ constant we can adopt a power-law dependence of the dissipation rate with radius of $d\mathcal{E}_B/dR=\mathcal{A}R^{-q}$, 
where $\mathcal{A}= \frac{\mathcal{E}_B}{\int_{R_{\rm begin}}^{R_{\rm end}}R^{-q}dr}$ is a normalization factor.

As the photons begin to escape the outflow and reconnection begins, we could expect CR acceleration
to occur \citep[e.g.,][]{Giannios10crmag,Kagan+15recon}, which is also facilitated by the decreasing
chance of scattering against photons. The cooling mechanisms for accelerated CRs include mainly
synchrotron and inverse-Compton (IC) losses, inelastic $pp$ scattering, Bethe-Heitler pair-production and photomeson production processes.

\subsection{Cooling timescales of protons}
\label{sec:cooling}
We consider the cooling at the diffusion radius, where we assume CR acceleration begins ($R_{\rm begin} \sim R_D$). The density there can be expressed as $\rho_D=\rho_0(R_{\rm cr}/R_0)^{-2}(R_D/R_{\rm cr})^{-3}$, where $\rho_0=1\,\mathrm{g\,cm^{-3}}, R_0=10^9~R_9\,\rm cm$ is the density and radius at the initial stage of the hot coronal outflow. Also, the magnetic field $B_D=B_0(R_D/R_0)^{-1}$ and blackbody temperature $T_D=T_0(R_{\rm cr}/R_0)^{-2/3}(R_D/R_{\rm cr})^{-1}$, and we choose nominal values $T_0=10^8~T_8\,\mathrm{K},\, B_0=10^{10}~B_{10}\,\rm G$ (see Section \ref{sec:outflow}). 
In this work, we calculate cooling rates of high-energy protons, using the numerical code developed in \citet{mur07} and \citet{mur08ph}.
\par
High-energy protons lose their energies through adiabatic, radiative and hadronic processes. The adiabatic cooling timescale $t_{\rm ad}$ is comparable to the dynamical timescale. 
Radiative cooling includes synchrotron and IC scattering, and the synchrotron cooling timescale is
\beq
t_{\rm syn}=\frac{6\pi m_p^4c^3}{\sigma_Tm_e^2B^2\epsilon_p}
\label{eq:sync}
\enq
and the IC cooling timescale is
\beq
t_{\rm IC}^{-1}=\frac{c}{2\gamma_p^2}\left(\frac{m_e^2}{m_p^2}\right)\pi r_e^2m_p^2c^4\int_0^\infty{\epsilon^{-2}\frac{dn}{d\epsilon}\frac{F(\epsilon,\gamma_p)}{\beta_p(\gamma_p-1)}d\epsilon},
\label{eq:IC}
\enq
where $\sigma_T$ is the
Thomson cross section, $\gamma_p$ is the Lorentz factor of protons and the expression of function $F(\epsilon,\gamma_p)$ can be found in \citet{mur07}. For a blackbody spectrum of scattering photons, the average photon energy is
$\epsilon=2.7kT$ and the average number density is $n\simeq19.232\pi\times\frac{1}{(hc)^3}\times(kT)^3$. Throughout the paper, we label the energies in the local frame as $\epsilon$ (e.g. $\epsilon_p,\,\epsilon_\nu$) and as $E$ in the observer's frame.

Hadronic cooling mechanisms mainly contain inelastic $pp$ collisions, the Bethe-Heitler and photomeson production processes in the following ways respectively,
$$p+p\longrightarrow p/n+N\pi$$
$$p+\gamma\longrightarrow p+e^{\pm},$$
$$p+\gamma\longrightarrow p/n+N\pi$$

We can expect muon neutrinos to be produced in $pp$ and $p\gamma$ processes, and electron neutrinos especially from muon decay.

The cooling timescale of inelastic $pp$ scattering is
\beq
t_{pp}=\frac{1}{c\sigma_{pp}n_p\kappa_{pp}}.
\label{eq:pp}
\enq
We estimate the proton number density as $n_p=\rho_DY_p/m_p\simeq 7.2\times 10^{11}{\,\rm cm}^{-3}$, where 
the assumed $Y_p=0.1$ is proton mass fraction of the ejected material since there will be heavy nuclei ejected from a carbon-oxygen white dwarf \citep{sch12, dan14}. The neutrino emission from these nuclei is expected in a low-energy range (see the Appendix) and is out of our interest. 
Assuming in each collision a fraction 50\% of the proton energy is lost and using the energy-dependent $pp$ cross section given by \citet{kel06}, we can get the $pp$ cooling timescale.

The photomeson production dominates the cooling at sufficiently high energies, and the timescale can be expressed as
\citep[e.g.,][]{ste68}
\beq
t_{p\gamma}^{-1}=\frac{c}{2\gamma_p^2}\int_{\bar{\epsilon}_{\rm th}}^\infty{d\bar{\epsilon}\sigma_{p\gamma}(\bar{\epsilon})\kappa_{p\gamma}(\bar{\epsilon})\bar{\epsilon}\int_{\bar{\epsilon}/2\gamma_p}^\infty{\epsilon^{-2}\frac{dn}{d\epsilon}d\epsilon}},
\label{eq:pgamma}
\enq
where $\bar{\epsilon}$ is the photon energy in the rest frame of proton, $\kappa_{p\gamma}$ is the inelasticity and $\bar{\epsilon}_{\rm th}$ is the threshold photon energy for the photomeson production process.

At relative higher energy, the protons start to cool through the BH pair-production process. 
The same formula is applied with the replacement of the cross section and threshold energy~\citep{cho92}. 
In particular, the high-energy BH cross section is $\sigma_{\rm BH}\approx(28/9)\alpha r_e^2\ln[(2\epsilon_p\epsilon)/(m_pm_ec^4)-106/9]$. 

The timescale of the magnetic reconnection acceleration is assumed to be comparable to that 
of diffusive shock acceleration, $t_{\rm acc}=\epsilon_p/(e\beta_{\mathrm{rec}}Bc)$, where 
$\beta_{\rm rec}$ is the reconnection speed, $\beta_{\rm rec}\sim0.1-0.2$ \citep[e.g.,][]{Giannios10crmag}. 

We now plot the inverse of all these timescales as functions of proton energy at $R_D$ in Figure 1.  The maximum proton energy can be found by equaling total energy loss time with acceleration time. Since we know that only the $pp$ reaction and photomeson production process will produce neutrinos, the other interactions would provide a strong suppression on the final neutrino spectrum. This suppression factor due to proton cooling can be written as
$\zeta_{\rm CRsup}$~\citep{mur08ph,wd09},
\beq
\zeta_{\rm CRsup}(\epsilon_\nu)=\frac{t_{pp}^{-1}+t_{p\gamma}^{-1}}{t_{\rm syn}^{-1}+t_{\rm IC}^{-1}+t_{\rm ad}^{-1}+t_{pp}^{-1}+t_{\rm BH}^{-1}+t_{p\gamma}^{-1}}.
\label{eq:suppr}
\enq

\begin{figure}
\plotone{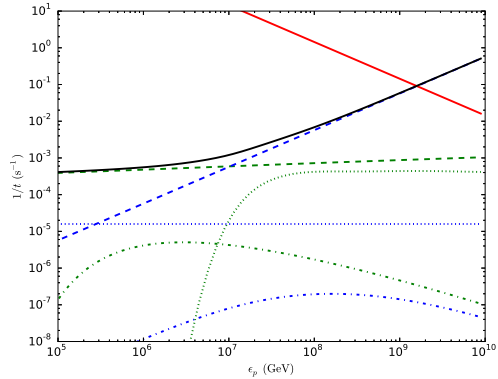}
\caption{The inverse of cooling timescales for protons at the diffusion radius: blue dashed-synchrotron, blue dotdashed-IC, blue dotted--adiabatic cooling, green dashed--inelastic $pp$ scattering, green dotdashed--BH process, green dotted--photomeson production, black solid--total. Also shown is the reconnection acceleration timescale--red solid.
\label{fig1}}
\end{figure}

Further on, the cooling of the mesons also need to be considered. It is similar to the proton radiative and hadronic cooling times are
\beq
t_{\rm syn}=\frac{6\pi m_\pi^4c^3}{\sigma_Tm_e^2B_r^2\epsilon_\pi},
\label{eq:pionsync}
\enq
\beq
t_{\rm had}=1/(c\sigma_{\pi p} n_p \kappa_{\pi p}).
\label{eq:pionhad}
\enq
The pion-proton scattering cross section is $\sigma_{\pi p}\approx 5\times10^{-26}\,\rm cm^2$ at the  energies of interest, and the inelasticity is $\kappa_{\pi p}=0.8$ \citep{Oli14}. For the emission region, we use the magnetic field $B_r$ that remains after reconnection events with the energy fraction $\epsilon_B=0.01$ \citep[e.g][]{med99}. For our outflow parameters, the meson goes from decay dominated to radiation cooling dominated. We can define the break energy for neutrinos, $\epsilon_{\nu,\rm brk}$ satisfies $t_{\rm dec}\equiv\gamma_\pi\tau_\pi\sim t_{\pi,\rm cool}$, and thus the suppression factor due to meson cooling is expressed to be
\beq
\zeta_{\pi \rm sup}(\epsilon_\nu)=\frac{t_{\rm dec}^{-1}}{t_{\rm dec}^{-1}+t_{\rm syn}^{-1}+t_{\rm had}^{-1}}
\label{eq:pionsup}
\enq
Since the mean lifetime of muons is much longer than that of pions, the break energy due to muon 
cooling is much lower than $\epsilon_{\nu,\rm brk}$, and all flavors from muon decay can contribute only 
at lower energies with no effect on the high-energy spectrum, so for simplicity we ignore the muon 
cooling effect.

\section{Neutrino Emission from Nearby Mergers}
\label{sec:indiv}
The neutrino spectrum should follow the initial proton spectrum if there are no 
energy-dependent suppression factors, such as Bethe-Heitler or radiative cooling losses.
We shall assume the initial accelerated proton spectrum to be a power-law 
$dN_p/d\epsilon_p\propto\epsilon_p^{-s}$ with index $s=2$, and we consider two suppression factors
which modify the neutrino spectrum, see eqs.(\ref{eq:suppr},\ref{eq:pionsup}).
The neutrino luminosity is then 
\begin{equation}
\epsilon_\nu L_{\epsilon_\nu}\propto \eta L_B\zeta_{\rm CRsup}(\epsilon_\nu)\zeta_{\pi \rm sup}(\epsilon_\nu).
\label{eq:nulum}
\end{equation}
For one single merger event with nominal parameters, we plot the neutrino spectrum in Figure 2. 
At the low energy end, the neutrino spectrum is flat, $\epsilon_\nu L_{\epsilon_\nu}\sim\rm const$. 
However, the neutrino flux has been suppressed by synchrotron, IC, adiabatic and BH cooling of protons. For higher energies,  the radiative cooling of pions become severe, leading to a strong suppression 
of the neutrino spectrum, and a sharp drop feature appears.

\begin{figure}
\plotone{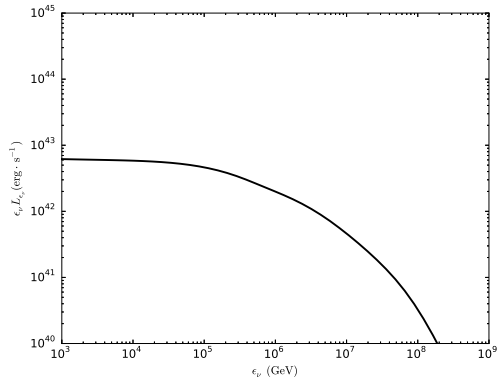}
\caption{The neutrino spectra of a single merger event. The total dissipated magnetic energy is assumed to be $\mathcal{E}_B=10^{50}\,\rm erg$.
\label{fig2}}
\end{figure}

Based on the neutrino spectrum above, it is of interest to discuss the prospects for the 
detection of individual nearby merger events, since the rate of these mergers is relatively high.
Let us consider a merger event at distance $D_L=10\,\rm Mpc$. 
We assume that the CR acceleration begins at $R_D$ and ends at roughly $R_{\rm end}\sim 100R_D$, and the CR efficiency is $\eta=0.1$. Therefore for our optimistic case of $\mathcal{E}_B=10^{50}\,\rm erg$, the total local injection power into CRs by white dwarf mergers is $Q_{\rm inj}=\eta\mathcal{R}\mathcal{E}_B\sim10^{45}\,\rm erg\,Mpc^{-3}\,yr^{-1}$, where the rate of white dwarf mergers is set to $\mathcal{R}\sim10^{-4}\,\rm Mpc^{-3}\,yr^{-1}$ \citep[e.g][]{bad12}, comparable to the rate of type Ia supernovae.

The neutrino fluence for single merger event can be expressed as 
\beq
\begin{split}
E_\nu^2\mathcal{F}_\nu(E_\nu)=&\int_{R_D}^{R_{\rm end}}dR\frac{K}{4(1+K)}\times \\
&\frac{d\mathcal{E}_{\rm CR}/dR}{4\pi D_L^2\ln{(E_{p,\rm max}/E_{p,\rm min})}}
\zeta_{\rm CRsup}(E_\nu)\zeta_{\pi \rm sup}(E_\nu),
\label{eq:fluence}
\end{split}
\enq
where $K$ denotes the average ratio of charged to neutral pions, with $K\approx1$ for $p\gamma$ and 
$K\approx2$ for $pp$ interactions \citep{mur16a}, the differential CR power $d\mathcal{E}_{\rm CR}/dR=\eta\mathcal{A}R^{-q}$ and 
we take $q=2$ in this section. The integration on $R$ gives $\mathcal{E}_{\mathrm {CR}}
\equiv\int_{R_D}^{R_{\mathrm {end}}}{\eta\mathcal{A}R^{-q}}=\eta \mathcal{E}_B\sim10^{49}\,\rm erg$
for the optimistic case.
Using the latest IceCube effective area $A(E_\nu)$ given by \citet{aar15c}, we can now estimate 
the number of neutrino events in the IceCube detector. The number of muon neutrinos above 1~TeV is
\beq
N(>1\mathrm{TeV})=\int_{1\mathrm{TeV}}^{E_{\nu,\max}}{dE_\nu A(E_\nu)\mathcal{F}_\nu(E_\nu)}.
\label{eq:num}
\enq
For our parameters, $N(>1\mathrm{TeV})\sim0.08$. Note that the fluence depends on the inverse of 
the distance square, therefore, for a possible IceCube observation, this source should be within 
$D_{L}^\prime=(0.08/1)^{1/2}\times10\,\rm Mpc\sim3\,Mpc$ from the earth. This implies that on average
we would have to wait for $((4/3)\pi D_{L}^{\prime3}\times\mathcal{R})^{-1}\sim 10^2\,
\mathcal{R}_{-4}^{-1}\eta_{-1}^{-1}~ \rm yr$. Poisson fluctuation effects or a higher rate or efficiency
(or including related mergers) could conceivably reduce this wait time.

If such merger events occur very close to us, one may ask whether {\it Fermi-LAT} can observe the 
GeV $\gamma$-ray signal from the source.  Importantly, this WD merger scenario is optically 
thick, and they may be considered as hidden sources \citep{mur16a}. 
The Thomson optical depth is $\tau_T=n_p\sigma_T R=\tau_T(R_D)(R/R_D)^{-2}$, where $\tau_T(R_D)=n_p\sigma_T R_D\sim30$ is the depth at the diffusion radius. Correspondingly, the photosphere radius is $R_{\rm ph}\sim6R_D$, so we can expect that most of the accompanying high-energy gamma-rays may be absorbed inside the fireball. More accurately, to verify this argument, we need to consider the $\gamma\gamma$ annihilation, for which the cross section is
\beq
\sigma(\epsilon_\gamma, \epsilon)=
\frac{\pi r_e^2}{2}(1-\beta^2)[2\beta(\beta^2-2)+(3-\beta^4)\ln(\frac{1+\beta}{1-\beta})],
\label{eq:sigmagg}
\enq
where $\beta=(1-\frac{m_e^2c^4}{\epsilon_\gamma\epsilon})^{1/2}$, classical electron radius $r_e=2.818\times10^{-13}\,\rm cm$, and the thermal photon energy $\epsilon=2.7kT=\epsilon(R)$,  density $n=n(R)$, so the $\gamma\gamma$ optical depth is $\tau_{\gamma\gamma}=\tau_{\gamma\gamma}(R)=n\sigma(\epsilon_\gamma, \epsilon)R$. 
For photons of energy below 1\,GeV, $\tau_{\gamma\gamma}<1$ so they may still escape from the source and could possibly trigger {\it Fermi-LAT}. At a distance $D_L=10\,\rm Mpc$, the accompanying 
fluence of GeV photons is of order $\mathcal{F}_\gamma\sim 8\times10^{-3}\,\rm GeV\cdot cm^{-2}$. 
The differential sensitivity of {\it Fermi-LAT} at 1\,GeV is about $\mathcal{F}_{\mathrm {sens}}\sim6.4\times10^{-11} 
\,\rm erg\cdot cm^{-2}\cdot s^{-1}$. For a merger event of duration $\mathcal{T}\sim 10^4\,\rm s$, at that distance it would be likely to be observed. From the non-detection of such merger events 
by LAT we can derive a constraint, namely $\frac{4}{3}\pi d_{\max}^3\times \mathcal{R}\times
\mathcal{T}_{\rm obs}<1$, where the {\it Fermi-LAT} operation time $\mathcal{T}_{\rm obs}\sim 8\,\rm yr$, the WD merger 
event rate is $\mathcal{R}=10^{-4}\,\rm Mpc^{-3}\cdot yr^{-1}$ and the maximum distance 
$d_{\max}$ can be related to the CR energy $\mathcal{E}_{\rm CR}$ as 
$\frac{\mathcal{E}_{\rm CR}}{d_{\max}^2\cdot \mathcal{F}_{\rm sens}}=\frac{10^{49}\rm erg}
{(10\mathrm {Mpc})^2\cdot \mathcal{F}_\gamma}$. This leads to an upper limit of $\mathcal{E}_{\rm CR}
\lesssim2.2\times10^{47}\,\rm erg$. Finally, substituting $\eta=0.1$ we can get a rough constraint 
on the dissipated magnetic energy $\mathcal{E}_B\lesssim 2\times10^{48}\,\rm erg$.

However, we neglected the matter attenuation effect in the above estimate, e.g., even GeV photons may undergo BH pair-production interactions and be strongly attenuated \citep{mur15}. We adopt the attenuation coefficients in \citet{mur15} and find that the optical depth is $\tau_M(R_D)\sim 7.7$ for $1\,\rm GeV$ photons. 
That means that the flux of GeV photons will fade away sufficiently (by a factor of $e^{-\tau_M}$) 
and would not trigger any detection.
If the ejecta is clumpy, some GeV gamma-rays may escape, which may lead to a detectable signal for {\it Fermi-LAT}. However, one should keep in mind that in practice the chance of being observed should be less due to the finite field of view and the response time of the instrument, or other accidental reasons. 
Most of the high-energy photons are absorbed and the re-radiation of electron-positron pairs is more likely to be concentrated in the soft X-ray band. Eventually, a significant fraction of the energy would be radiated in the ultraviolet or optical band.  Using the diffusion time $t_{\rm diff}\approx \tau_T R/c\sim1.3\times{10}^6$~s, the luminosity of this transient is estimated to be $L_{\gamma}\approx{\mathcal E}_B/t_{\rm diff}\sim10^{42}-{10}^{44}\,\rm erg\,s^{-1}$ or $L_\gamma=4\pi R_D^2ca_RT_D^4\sim3\times10^{42}\,\rm erg\,s^{-1}$, which is 
an order of magnitude lower than the peak luminosity of nearby event SN 2011fe ($a \,few\times10^{43}\,\rm erg\,s^{-1}$) but detectable. Thus, the non-detection of bright thermal transients in the optical survey would also enable us to put constraints on the WD merger model especially if the non-thermal injection is large.

\section{Diffuse neutrino flux}
\label{sec:difnuflux}
For a flat CR energy spectrum, the local neutrino energy budget is estimated to be
\beq
\begin{split}
\epsilon_\nu Q_{\epsilon_{\nu_i}}=&\int_{R_D}^{R_{\rm end}}\frac{K}{4(1+K)}\times\\
&\frac{\eta\mathcal{R}\mathcal{A}R^{-q}dR}{\ln{(\epsilon_{p,\rm max}/\epsilon_{p,\rm min})}}\zeta_{\rm CRsup}(\epsilon_\nu)\zeta_{\pi\rm sup}(\epsilon_\nu).
\label{eq:localnu}
\end{split}
\enq
Defining $Q_\nu$ in the comoving volume, the diffuse neutrino flux per flavor is given by \citep[e.g.,][]{mur16a}
\beq
E_{\nu}^2\Phi_{\nu_i}=\frac{c}{4\pi H_0}\int_0^{z_{\rm max}}\frac{\epsilon_\nu Q_{\epsilon_{\nu_i}}S(z)}{(1+z)^2\sqrt{\Omega_M(1+z)^3+\Omega_\Lambda}}dz,
\label{eq:diffnu}
\enq
where we assume that the source evolution traces the cosmological star formation history, which can be 
expressed as $S(z)=[(1+z)^{-34}+(\frac{1+z}{5000})^3+(\frac{1+z}{9})^{35}]^{-0.1}$ \citep{hop06,yuk08},  
$z_{\rm max}=4$, and the cosmology parameters are $H_0=67.8\,{\rm km\,s^{-1}\,Mpc^{-1}}, \Omega_M=0.308$ 
\citep{pla15}.

Figure 3 shows the diffuse neutrino flux of our model. The black solid line is for the fiducial values 
$Q_{\rm inj}=\eta\mathcal{R}\mathcal{E}_B\sim10^{43}\,\rm erg\,Mpc^{-3}\,yr^{-1}$ based on 
$\mathcal{E}_B=10^{48}~\rm erg$, while the dashed line is for the optimistic case $Q_{\rm inj}=10^{45}\,
\rm erg\,Mpc^{-3}\,yr^{-1}$ corresponding to $\mathcal{E}_B=10^{50}~\rm erg$.   
In the fiducial CR injection scenario, the WD mergers does not contribute much to the diffuse neutrino flux, but the CR injection rate of the WD mergers is highly uncertain.  We can see that, for our optimistic injection scenario, this kind of WD merger events is a potentially interesting source to account for the IceCube neutrino data. However, a very large injection power $Q_{\rm inj}\sim 8\times10^{45}\,\rm erg\,Mpc^{-3}\,yr^{-1}$ (dotted line in Figure 3) is needed to reach the flux level at $\sim30\,\rm TeV$, which is unlikely from WD-WD mergers alone, although such numbers might be obtainable if one adds the combined effect of similar merger events such as WD-NS and NS-NS mergers \citep[e.g][]{met12}. 
\begin{figure}[t]
\plotone{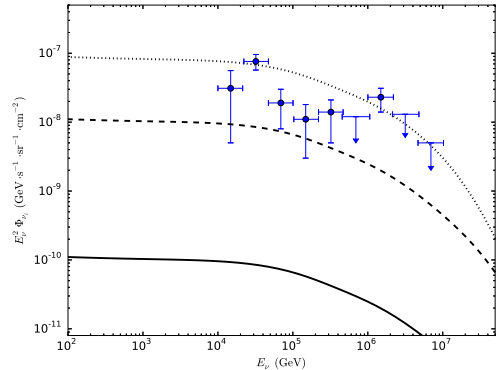}
\caption{Diffuse neutrino flux of the WD merger scenario. The solid line represent a nominal CR injection power $Q_{\rm inj}=\eta\mathcal{R}\mathcal{E}_B=(0.1)\times(10^{-4}\,\rm Mpc^{-3}\,yr^{-1})\times(10^{48}\,erg)=10^{43}\,\rm erg\,Mpc^{-3}\,yr^{-1}$ and the dashed line shows an optimistic case of 100 times higher ($\mathcal{E}_B=10^{50}\,\rm erg$). For illustration, we also plot a case in which the injection power needed to reach the IceCube data around $30\,\rm TeV$ is $Q_{\rm inj}\sim8\times10^{45}\,\rm erg\,Mpc^{-3}\,yr^{-1}$ (dotted line). The IceCube data is indicated by blue points \citep{aar15}. Here we take $q=2$ as an example. 
\label{fig3}}
\vspace{-1.\baselineskip}
\end{figure}

In Figure 4 we investigate the effects of varying the parameter $q$ on the magnetic reconnection rate 
and the final spectrum, taking the optimistic case $\mathcal{E}_B=10^{50}\,\rm erg$ as an example. 
We see that the diffuse neutrino flux depends strongly on the magnetic dissipation rate. For larger $q$, the neutrino flux is higher. This is easy to see if we take into account the radial dependence that $t_{pp}^{-1}\propto n_p\propto R^{-3},\,t_{p\gamma}^{-1}\propto n\propto R^{-3}$, and $t_{\rm sync}^{-1}\propto R^{-2},\,t_{\rm ad}^{-1}\propto R^{-1}$. From eq.(\ref{eq:suppr}) we can derive that $\zeta_{\rm CRsup}$ decreases with $R$. If the magnetic energy dissipates more quickly (larger $q$), the shape of the spectrum will be closer to the shape at $R_D$, where $\zeta_{\rm CRsup}$ is larger and thus leads to higher neutrino flux.
\begin{figure}
\plotone{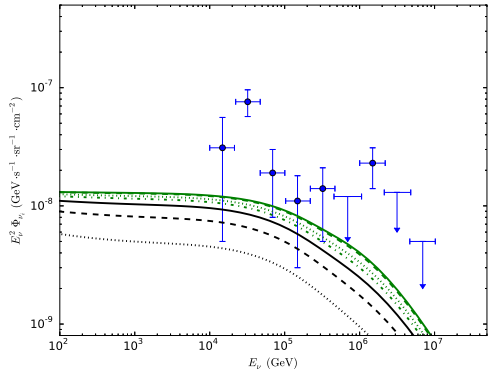}
\caption{Same as Figure 3 but with different $q$. Green solid: $q=10$; green dashed: $q=5$; green dotted: $q=3$; green dotdashed: $q=2.5$; black solid: $q=2.0$; black dashed: $q=1.5$; black dotted: $q=1.0$. The injection power is set to $Q_{\rm inj}=(0.1)\times(10^{-4}\,\rm Mpc^{-3}\,yr^{-1})\times(10^{50}\,erg)=10^{45}\,\rm erg\,Mpc^{-3}\,yr^{-1}$.
\label{fig4}}
\end{figure}

The neutrino and gamma-ray energy generation rates are conservatively related as 
$\epsilon_{\gamma}^2\Phi_{\gamma}=\frac{4}{K}\epsilon_{\nu}^2\Phi_{\nu}|_{\epsilon_\nu=0.5\epsilon_\gamma}$. 
The diffuse gamma-ray flux is shown by the red lines in Figure 5 (Only the component attenuated by 
$\gamma\gamma$ absorption is relevant, since cascades occur in the synchrotron-dominated regime).  
We can clearly see that even for the very optimistic case of an injection power $8\times10^{45}\,\rm erg\,Mpc^{-3}\,yr^{-1}$, the diffuse gamma-ray flux from these sources is below the extragalactic gamma-ray background measured by {\it Fermi-LAT}.

\begin{figure}
\plotone{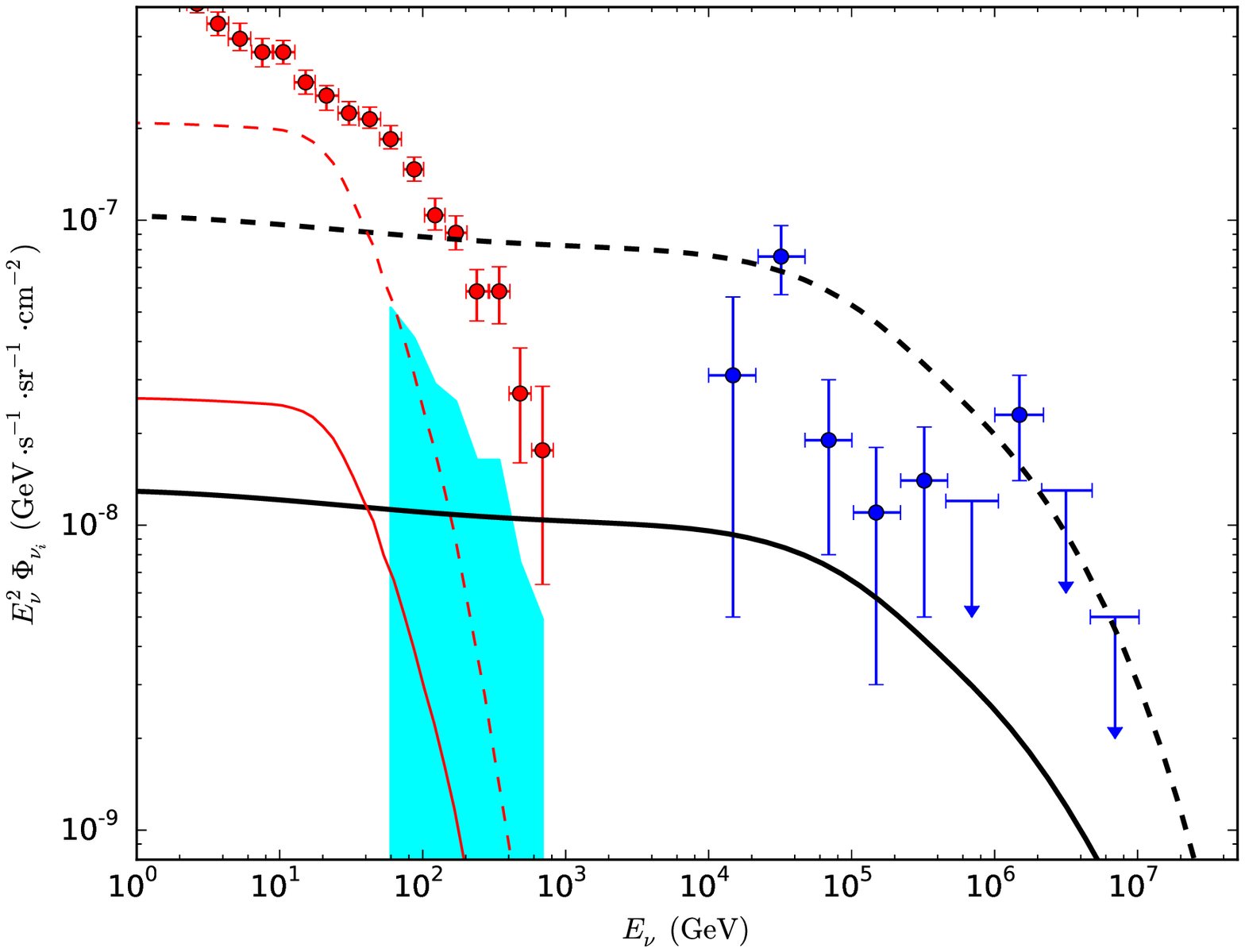}
\caption{The ``optimistic'' diffuse gamma-ray flux of the WD merger scenario, which shows that the predicted gamma-ray flux should be far below the extragalactic gamma-ray background measured by {\it Fermi-LAT} (red data points) \citep{ack15}. The cyan area shows the allowed region for the non-blazar gamma-ray flux in \citet{dim16}. Only the $\gamma\gamma$ absorption effect is included in this figure, although the Bethe-Heitler pair production in the ejecta is also likely to be important. The thin red solid line is for the optimistic injection with $Q_{\rm inj}=10^{45}\,\rm erg\,Mpc^{-3}\,yr^{-1}$ and thin red dashed line is for the even more optimistic case of 8 times higher. The thick black lines are the neutrino fluxes, correspondingly. We take $q=2$ as an example.
\label{fig5}}
\end{figure}

\section{Discussions and Conclusions}
\label{sec:disc}
In this work we discussed the high-energy neutrino emission from WD mergers, and showed that they may be potentially interesting sources for IceCube observations.  
Since the total CR injection power is highly uncertain, we considered both a nominal case of $\mathcal{E}_B=10^{48}\,\rm erg$ and an optimistic case of $\mathcal{E}_B=10^{50}\,\rm erg$. 
This kind of merger events at cosmological distances are essentially hidden sources and would not contribute significantly to the high energy gamma-ray background, thus eliminating the tension that exists with {\it Fermi-LAT} for both the (optically thin) hadronuclear and photohadronic scenarios. Note that the $p\gamma$ efficiency is larger if $\mathcal{E}_B$ is sufficiently larger than $10^{48}\,\rm erg$, since the synchrotron photons due to reconnections may become dominant as target photons, which is beyond the scope of this work but interesting to investigate in future.
Besides, we found that the diffuse neutrino flux depends strongly on the dissipation rate of magnetic energy. 
The faster the magnetic energy dissipates, the more high-energy neutrino flux is expected.

Even if these mergers are not responsible for IceCube's diffuse neutrino flux, searches for 
high-energy neutrino and gamma-ray signals from a single merger event are useful and interesting.  
The neutrino spectra of a single merger event is characterized by a global suppression which 
is caused by the other cooling mechanisms competing with $pp$ and $p\gamma$ neutrino production process, and a sharp drop at energies $\gtrsim100\,\rm TeV$ due to the radiative cooling suppression of pions.
For individual WD mergers at Mpc scale distances, the attenuation of the gamma-rays by $\gamma\gamma$ absorption below $1\,\rm GeV$ is negligible, allowing a constraint to be placed on such mergers by the fact that they have not been observed by {\it Fermi-LAT}. 
Based on a simple estimate, this constraint implies a total CR energy production per event of $\lesssim10^{47}$ erg, which in turn constrains the contribution that such sources can make to the diffuse neutrino background. 
However, this constraint should be relieved if we take into account the matter attenuation effect, 
which is likely to occur in a non-clumpy and spherical setup.

It is possible that not all WD mergers result in the type of remnants discussed here. Some WD merger 
events may be able to ignite a thermonuclear explosion promptly after merger \citep{1984ApJ...277..355W,1984ApJS...54..335I,dan12}, and these will be identified as 
type Ia supernovae (e.g., if mergers are violent) \citep{2010Natur.463...61P,2012ApJ...747L..10P,liu16}.  There are also other possible outcomes of WD mergers, such as a massive fast-rotating WD \citep[e.g.,][]{seg97, gar12} or an accretion-induced collapse into a NS \citep[e.g.,][]{sai85, sai04}.  In this work we have mainly focused on the merger events that do not explode on a dynamical time \citep{ras09}, yet no such events have been identified so far. Nevertheless, with accumulating {\it Fermi-LAT} and IceCube operation hours, this kind of merger events may finally be detected, allowing a test of our model in the near future.

\acknowledgements
We acknowledge support by the National Basic Research Program of China (973 Program grant 2014CB845800 and the National Natural Science Foundation of China grant 11573014 (D.X. and D.Z.G.), by the program for studying abroad supported by China Scholarship Council (D.X.), by Pennsylvania State University (K.M.) and by NASA NNX13AH50G (P.M.). The work of K. M. is also supported by NSF Grant No. PHY-1620777.

\clearpage
\appendix
\section*{CALCULATION OF THE MAXIMUM ENERGY OF NUCLEI}
The ejected material of WD merger remnant contains nuclei like carbon and oxygen. Let us consider nuclei with average mass number $A\sim13$ and atomic number $Z\sim7$, so the number density at $R_D$ is $n_A=\rho_DY_A/Am_p\simeq5.0\times10^{11}{\,\rm cm}^{-3}$. These nuclei can  undergo inelastic hadronuclear scattering with protons and with themselves (spallation). The dependence of the 
proton-nucleus cross section on mass number is approximated to be $\sigma_{pA}\propto A^{2/3}$ \citep{let83}. 
At the diffusion radius, the timescale is $t_{pA}\sim (n_A\sigma_{pA}c)^{-1}\simeq3\times10^{2}\,\rm s$, and the
spallation timescale is of the same order, the cross section being a factor of $\sim2$ larger. Also, for 
high energy nuclei (with Lorentz factor $\gamma_A>10^6$), the energy of the ambient thermal photons is boosted into the MeV range in the rest frame of the nucleus, where the giant dipole resonance (GDR) 
leads to photodisintegration of the nuclei \citep{anc08}. The timescale for this process is estimated to be
$t_{\rm GDR}\sim (n_{ph} \sigma_{\rm GDR}c)^{-1}\sim6.8\,\rm s$, even shorter than $t_{pA}$. Moreover, the photodisintegration optical depth at $R_D$ is $\tau_{\rm GDR}= n_{ph} \sigma_{\rm GDR}R\gg 1$, so that the CR nuclei whose energies exceed the GDR threshold are disintegrated into protons and neutrons. The enrichment of the proton outflow amounts to a factor of order unity. These protons will be reaccelerated to high energies before producing neutrinos, and for simplicity here we have neglected this effect. The maximum energy of nuclei is determined by $t_{A,\rm acc}^{-1}=t_{pA}^{-1}+t_{AA}^{-1}+t_{\rm GDR}^{-1}$, and we plot in Figure 6 the maximum Lorentz factor of protons and nuclei. Throughout the whole expansion process we have $\gamma_{A,\max}<\gamma_{p,\max}$. Further, we need to check whether the requirement of GDR ($\gamma_AT>\,\rm MeV$) is always satisfied. Since the decrease of the thermal photon energy is $T\propto R^{-1}$ and the numerical solution shows that $\gamma_{A,\max}$ grows more slowly than $\propto R$, the product of $\gamma_A \times T$ decreases with $R$. At the end stage $R_{\rm end}\sim100R_D$, we have $\gamma_{A,\rm end}\times T(R_{\rm end})\sim 240\,\rm MeV$ thus an effective GDR photodissociation is always expected.

\begin{figure}
\begin{center}
\includegraphics[width=3.00in]{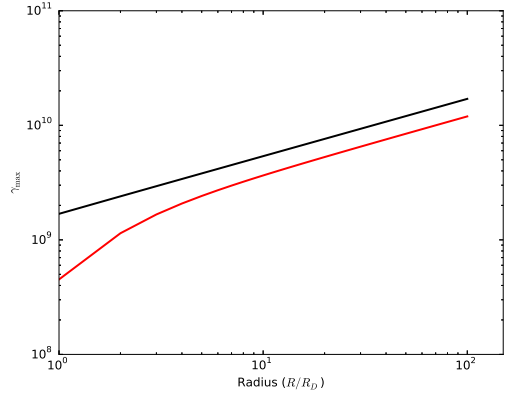}
\caption{The maximum Lorentz factor of protons and nuclei, depending on radius. Black solid line represents protons ($\gamma_{p,\max}$) and red solid line is for nuclei ($\gamma_{A,\max}$).
\label{fig6}}
\end{center}
\end{figure}

\par

\clearpage


\clearpage


\begin{thebibliography}{}
\bibitem[Aartsen et al.(2013a)]{aar13a}Aartsen, M. G. \& IceCube Collaboration 2013, \prl, 111, 1103
\bibitem[Aartsen et al.(2013b)]{aar13b}Aartsen, M. G. \& IceCube Collaboration 2013, Science, 342, 6161
\bibitem[Aartsen et al.(2015a)]{aar15}Aartsen, M. G. \& IceCube Collaboration 2015, \apj, 809, 98
\bibitem[Aartsen et al.(2015b)]{IC3+15tevnu}Aartsen, M. G. \& IceCube Collaboration 2015, \prd, 91, 022001
\bibitem[Aartsen et al.(2015c)]{aar15c}Aartsen, M. G. \& IceCube Collaboration 2015, {arxiv:1510.05222v3}
\bibitem[Aartsen et al.(2016)]{aar16}Aartsen, M. G. \& IceCube Collaboration 2016, arXiv:1607.08006 
\bibitem[Ackermann et al.(2015)]{ack15}Ackermann, M., Ajello, M., Albert, A., et al. 2015, \apj, 799, 86
\bibitem[Ahlers \& Halzen(2015)]{Ahlers+15nurev}Ahlers, M., \& Halzen, F. 2015, {\it Rep. Prog. Phys.}, 78, 126901
\bibitem[Anchorodoqui et al.(2008)]{anc08}Anchordoqui, L. A., Hooper, D., Sarlar, S., \& Taylor, A. M. 2008, Aph, 29, 1
\bibitem[Anchordoqui et al.(2014)]{anc14}Anchordoqui, L.~A., Barger, V., Cholis, I., et al.\ 2014, Journal of High Energy Astrophysics, 1, 1
\bibitem[Badenes \& Maoz(2012)]{bad12}Badenes, C., \& Maoz, D. 2012, \apj, 794, L11
\bibitem[Beloborodov(2014)]{bel14}Beloborodov, A. M. 2014, \mnras, 438, 169
\bibitem[Belyaev et al.(2013)]{bel13}Belyaev, M. A., Rafikov, R. R., \& Stone, J. M. 2013, \apj, 770, 68
\bibitem[Bogomazov \& Tutukov(2009)]{bog09}Bogomazov, A. I., \& Tutukov, A. V. 2009, Astron. Rep., 53, 214
\bibitem[Chang \& Wang(2014)]{cha14}Chang, X. C., \& Wang, X. Y. 2014, \apj, 793, 131
\bibitem[Chang \& Wang(2015)]{cha15}Chang, X. C., \& Wang, X. Y. 2015, \apj, 805, 95
\bibitem[Chakraborty \& Izaguirre(2015)]{sov15}Chakraborty, S., \& Izaguirre, I. 2015, Phys. Lett. B., 745, 35
\bibitem[Chodorowski et al.(1992)]{cho92}Chodorowski, M. J., Zdziarski, A. A., \& Sikora, M. 1992, \apj, 400, 181
\bibitem[Dan et al.(2012)]{dan12}Dan, M., Rosswog, S., Guillochon, J., \& Ramirez-Ruiz E. 2012, \mnras, 422, 2417
\bibitem[Dan et al.(2014)]{dan14}Dan, M., Rosswog, S., Br\"{u}ggen, M., \& Podsiadlowski, P. 2014, \mnras, 438, 14
\bibitem[Fermi Collaboration(2016)]{dim16}Fermi-LAT Collaboration, 2016, {arXiv:1601.04323}.
\bibitem[Garc\'ia-Berro et al.(2012)]{gar12}Garc\'ia-Berro, E., Lor\'en-Aguilar, P., Aznar-Sigu\'an, G., et al. 2012, \apj, 749, 25
\bibitem[Giannios(2010)]{Giannios10crmag}Giannios, D. 2010, \mnras, 408, 46
\bibitem[Hopkins \& Beacom(2006)]{hop06}Hopkins, A. M., \& Beacom, J. F. 2006, \apj, 651, 142
\bibitem[Hoshino(2013)]{hos13}Hoshino, M. 2013, \apj, 773, 118
\bibitem[Iben \& Tutukov(1984)]{1984ApJS...54..335I} Iben, I., Jr., \& Tutukov, A.~V.\ 1984, \apjs, 54, 335
\bibitem[Ji et al.(2013)]{ji13}Ji, S. Q., Fisher, R. T., Garc\'ia-Berro, E., et al. 2013, \apj, 773, 136
\bibitem[Kagan et al.(2015)]{Kagan+15recon}Kagan, D., Sironi, L., Cerutti, B., \& Giannios, D. 2015, \ssr, 191, 545
\bibitem[Kashyap et al.(2015)]{kas15}Kashyap, R., Fisher, R., Garc\'ia-Berro, E., et al. 2015, \apj, 800, 7
\bibitem[Kelner et al.(2006)]{kel06}Kelner, S. R., Aharonian, F. A., \& Bugayov, V. V. 2006, \prd, 74, 034018
\bibitem[Kotera et al.(2009)]{kot09}Kotera, K., Allard, D., Murase, K., et al.\ 2009, \apj, 707, 370 
\bibitem[Letaw et al.(1983)]{let83}Letaw, J. R., Silberberg, R., \& Tsao, C. H. 1983, \apjs, 51, 271
\bibitem[Liu et al.(2016)]{liu16}Liu, D. D., Wang, B., Podsiadlowski, P., \& Han, Z. W. 2016, {arXiv:1607.00161}
\bibitem[Loeb \& Waxman(2006)]{loe06}Loeb, A., \& Waxman, E. 2006, \jcap, 5, 3
\bibitem[Medvedev \& Loeb(1999)]{med99}Medvedev, M. V., \& Loeb, A. 1999, \apj, 526, 697
\bibitem[M{\'e}sz{\'a}ros(2014)]{mes14} M{\'e}sz{\'a}ros, P.\ 2014, Nuclear Physics B Proceedings Supplements, 256, 241
\bibitem[M\'esz\'aros \& Waxman (2001)]{mw01}M\'esz\'aros, P., \& Waxman, E. 2001, \prl, 87, 171102
\bibitem[Metzger(2012)]{met12}Metzger, B. D. 2012, \mnras, 419, 827
\bibitem[Murase(2007)]{mur07}Murase, K. 2007, \prd, 76, 123001
\bibitem[Murase(2008)]{mur08ph}Murase, K. 2008, \prd, 78, 101302(R)
\bibitem[Murase et al.(2008)]{mur08}Murase, K., Inoue, S., \& Nagataki, S. 2008, \apj, 689, L105
\bibitem[Murase \& Ioka(2013)]{mi13}Murase, K., \& Ioka, K. 2013, \prl, 111, 121102
\bibitem[Murase \& Waxman(2016)]{mw16}Murase, K., \& Waxman, E. 2016, arXiv:1607.01601.
\bibitem[Murase et al.(2013)]{mur13}Murase, K., Ahlers, M., \& Lacki, B. C. 2013, \prd, 88, 121301(R)
\bibitem[Murase et al.(2016a)]{mur16a}Murase, K., Guetta, D., \& Ahlers, M. 2016, \prl, 116, 071101
\bibitem[Murase et al.(2016b)]{mur16b}Murase, K., Kashiyama, K., M\'esz\'aros, P., Shoemaker, I., \& Senno, N. 2016., \apj, 822, L9
\bibitem[Murase et al.(2014)]{mur14}Murase, K., Inoue, Y., \& Dermer, C. D. 2014, \prd, 90, 023007
\bibitem[Murase et al.(2015)]{mur15}Murase, K., Kashiyama, K., Kiuchi, K., \& Bartos, I. 2015, \apj, 805, 82
\bibitem[Nelemans et al.(2001)]{nel01}Nelemans, G., Yungelson, L. R., Portegies Zwart, S. F., \& Verbunt, F. 2001, \aap, 365, 491
\bibitem[Olive et al.(2014)]{Oli14}Olive, K. A., et al. (Particle Data Group) 2014, Chin. Phys. C. 38, 090001
\bibitem[Pakmor et al.(2010)]{2010Natur.463...61P} Pakmor, R., Kromer, M., R{\"o}pke, F.~K., et al.\ 2010, \nat, 463, 61 
\bibitem[Pakmor et al.(2012)]{2012ApJ...747L..10P} Pakmor, R., Kromer, M., Taubenberger, S., et al.\ 2012, \apjl, 747, L10
\bibitem[Plank Collaboration(2015)]{pla15}Planck Collaboration 2015, {arXiv:1502.01589v2}
\bibitem[Raskin et al.(2009)]{ras09}Raskin, C., Scannapieco, E., Rhoads, J., \& Della Valle, M. 2009, \apj, 707, 74
\bibitem[Saio \& Nomoto(1985)]{sai85}Saio, H., \& Nomoto, K. 1985, \aap, 150, L21
\bibitem[Saio \& Nomoto(2004)]{sai04}Saio, H., \& Nomoto, K. 2004, \apj, 615, 444
\bibitem[Schwab et al.(2012)]{sch12}Schwab, J., Shen, K. J., Quataert, E., Dan, M., \& Rosswog, S. 2012, \mnras, 427, 190
\bibitem[Segretain et al.(1997)]{seg97}Segretain, L., Chabrier, G., \& Mochkovitch, R. 1997, \apj, 481, 355
\bibitem[Senno et al.(2015)]{nick15}Senno, N., M\'esz\'aros, P., Murase, K., Baerwald, P., \& Rees, M. J. 2015, \apj, 806, 24
\bibitem[Senno et al.(2016)]{senno16}Senno, N., Murase, K., \& M\'esz\'aros, P. 2016, \prd, 93, 083003
\bibitem[Stecker(1968)]{ste68}Stecker, F.~W.\ 1968, \prl, 21, 1016 
\bibitem[Tamborra et al.(2014)]{tam14}Tamborra, I., Ando, S., \& Murase, K. 2014, \jcap, 9, 43
\bibitem[Tamborra \& Ando(2016)]{ta16}Tamborra, I., \& Ando, S. 2016, \prd, 93, 053010
\bibitem[Uzdensky(2011)]{Uzdensky11magrec}Uzdensky, D. A. 2011, \ssr, 160, 45
\bibitem[Wang \& Loeb(2016)]{wl16}Wang, X., \& Loeb, A.\ 2016, arXiv:1607.06476
\bibitem[Wang \& Dai(2009)]{wd09}Wang, X. Y., \& Dai, Z. G. 2009, \apj, 691, 67
\bibitem[Waxman(2013)]{wax13}Waxman, E.\ 2013, arXiv:1312.0558 
\bibitem[Webbink(1984)]{1984ApJ...277..355W} Webbink, R.~F.\ 1984, \apj, 277, 355
\bibitem[Xiao \& Dai(2014)]{xia14}Xiao, D., \& Dai, Z. G. 2014, \apj, 790, 59
\bibitem[Xiao \& Dai(2015)]{xia15}Xiao, D., \& Dai, Z. G. 2015, \apj, 805, 137
\bibitem[Xiao et al.(2016)]{xia16}Xiao, D., M\'esz\'aros, P., Murase, K., \& Dai, Z. G. 2016, \apj, 826, 133
\bibitem[Y\"{u}ksel et al.(2008)]{yuk08}Y\"{u}ksel, H., Kistler, M. D., Beacom, J. F., \& Hopkins, A. M. 2008, \apj, 683, L5
\bibitem[Zhu et al.(2015)]{zhu15}Zhu, C. C., Pakmor, R., Van Kerkwijk, M. H., \& Chang, P. 2015, \apj, 806, L1 


\end{thebibliography}
\end{document}